\documentclass[a4paper,10pt]{article}

\title{ Noncommutative  Quantum Gravity }
\author{Mir Faizal\\
Mathematical Institute,  University of Oxford, \\
Oxford
OX1 3LB}
\date{}
\begin{document}

\maketitle

\begin{abstract}
We discuss the BRST and anti-BRST symmetries  for perturbative quantum 
gravity in noncommutative spacetime. 
In this noncommutative  perturbative quantum gravity the sum
 of the classical Lagrangian density with a gauge fixing term
 and a ghost term  is shown to be invariant the noncommutative  
BRST and the noncommutative anti-BRST transformations. 
We analyse the gauge fixing term and the ghost term in both linear
 as well as non-linear gauges. 
 We also discuss the unitarity evolution of the theory and
analyse the violation of unitarity of by   
introduction of a bare mass term in the  noncommutative BRST and the 
noncommutative anti-BRST transformations.
\end{abstract}



\section{Introduction}
Noncommutative spacetime was  originally
studied as a mechanism for providing a natural cut to control ultraviolet
divergences \cite{1}. However this motivation to study noncommutative field theory 
 ended with  the success of the renormalization  theory. The real motivation behind the 
study of  noncommutative  field theory is its close  relation  to the string theory \cite{2as}.
  The presence of an antisymmetric tensor background
along the $D$-brane  world volumes  is an important
source for noncommutativity in string theory \cite{2,3}.

The relation between noncommutative field theories and quantum gravity has been 
studied by many authors \cite{21,22,23, 23aa}.
The relation between noncommutative  gravity and the cosmological constant
 has been examined \cite{24}. In doing so 
it was concluded that noncommutative gravity leads to the existence 
of a tiny non zero cosmological constant of the order of the square of the Hubble constant. 
 Black holes 
in noncommutative gravity has also been thoroughly studied \cite{25, 26, 27}. 
Even in noncommutative gravity there is  no correlation between the different 
modes of radiation and so information does not come out continuously during 
the evaporation process. However, due to spacetime noncommutativity, 
information might be preserved by a stable black hole remnant \cite{28}.

Perturbative quantum gravity on noncommutative flat spacetime has also been discussed \cite{29}. 
In this paper will will discuss the 
BRST and anti-BRST symmetries for this noncommutative perturbative quantum gravity. 
It may be be remarked that the BRST symmetry 
for noncommutative Yang-Mills theories has already been analysed \cite{30, 31, 32, 33a}.
The BRST symmetry for spontaneously broken gauge theories in noncommutative spacetime
 has also been discussed \cite{33b}.
In case of noncommutative gauge theory 
the Hilbert space of physical states is determined by the cohomology space of 
the BRST operator as in the commutative case. 
The anti-BRST symmetry for noncommutative
Yang-Mills theories has also been analysed \cite{34}.

The BRST and the anti-BRST symmetries for commutative  
perturbative quantum gravity in flat spacetime
 have been studied by a number of  authors \cite{4,5,6} 
and their work has been summarized by N. Nakanishi and I. Ojima \cite{7}.
 The BRST symmetry in two dimensional curved  spacetime has 
 been thoroughly studied  \cite{8,9,10}.  The BRST and the anti-BRST symmetries 
for topological quantum gravity in  curved spacetime  have also been studied  \cite{11, 12}.
 All this work has been done in linear gauges. 
The BRST and anti-BRST symmetries for perturbative quantum gravity 
in non-linear gauges has also been analysed \cite{1a}. 
This analysis was done in arbitrary dimensional curved spacetime.

In this paper we shall study the BRST and the anti-BRST symmetries for
 noncommutative perturbative quantum gravity 
in both linear and non-linear gauges. 
Furthermore, it will be demonstrated that the addition of a bare mass 
term  violates   the nilpotency of the noncommutative BRST and the noncommutative anti-BRST 
transformations and this in turn violates  unitarity of the
 resultant theory. 
  
\section{Linear Gauges}
In perturbative gravity one splits the full metric $g_{ab}^{(f)}$ into the metric for the
 background spacetime  and a small perturbation around it. 
The covariant derivatives along with the lowering and raising of indices are compatible with the 
metric for the background spacetime. 
The small perturbation is viewed as the field that is to be quantized. 
For simplicity we shall take our background spacetime to be flat. 
Then noncommutativity is introduced by replacing the all the
 fields by noncommutative fields and all the product of fields by 
the Moyal $*$-product. 
The Moyal $*$-product of two noncommutative  
fields say $h^{ab}$ and $h_{ab}$ is defined as follows \cite{2as}
\begin{equation}
 h^{ab}*h_{ab} (x) = \exp \left( \frac{i}{2} \theta^{ab} 
\partial_a \partial_b\right) h^{ab}(x) h_{ab}(x+\epsilon)|_{\epsilon =0}.
\end{equation}
The Lagrangian density for pure  gravity  on noncommutative spacetime  
is given by \cite{29}
\begin{equation}
 \mathcal{L}_c = \sqrt{g}^{(f)} * R^{(f)},
\end{equation}
where we have adopted  units such that $ 16 \pi G = 1$.
Here $R^{(f)} = g^{ab} * R^{(f)}_{ab}$, where 
\begin{equation}
 R^{(f)}_{ab} = {R^{c(f)} }_{acb} = 
\partial_{c}{\Gamma^{c(f)}_{bc}} 
- \partial_{b}\Gamma^{c(f)}_{ca} + 
\Gamma^{c(f)}_{cd} *\Gamma^{d(f)}_{ba}
 - \Gamma^{c(f)}_{bd}*\Gamma^{d(f)}_{ca}.
\end{equation}
Here $R^{d(f)}_{abc}$ statisfies the Bianchi indentity 
\begin{equation}
 \partial_e R^{d(f)}_{abc} + \partial_c  R^{d(f)}_{aeb}
 + \partial_b  R^{d (f)}_{ace}=0.
\end{equation}

We can expand $g^{(f)}_{ab}$ as follows
\begin{equation}
 g^{(f)}_{ab} = \eta_{ab} + h_{ab}.
\end{equation}
The expansion of the Ricci scalar $R^{(f)}$ and the  metric $g^{(f)}_{ab}$ in terms of 
  the Minkowski metric  $\eta_{ab}$ and a small  perturbation around it $h_{ab}$ generates the Lagrangian density  for noncommutative
perturbative quantum gravity. There are  infinitely many terms in the Lagrangian
 for this noncommutative perturbative quantum gravity. 

All the degrees of freedom in $h_{ab}$ are not physical. This
is because the  Lagrangian density for it is invariant under a gauge transformation, 
\begin{equation}
\delta_\Lambda h_{ab} = \partial_a \Lambda_b + \partial_b \Lambda_a + \pounds_{(\Lambda)} h_{ab}, 
\end{equation}
where
\begin{equation}
\pounds_{(\Lambda)} h_{ab} = \Lambda^c *\partial_c h_{ab} + h_{ac}*\partial_b \Lambda^c + h_{cb}* \partial_a \Lambda^c.
\end{equation}
 
These unphysical degrees of freedom will give rise to constraints \cite{17} in the canonical
 quantization  and divergences in the partition function \cite{18} in the path integral quantization.
 So before we can quantize this theory, we need to remove these unphysical degrees of freedom. This can 
be done by the addition of a noncommutative gauge fixing term and a noncommutative ghost term. The sum of 
a gauge fixing term and a ghost term can now be written as  
\begin{eqnarray}
\mathcal{L}_g &=& -\frac{i}{2}s \overline{s} (h^{ab}*h_{ab})+ \frac{i\alpha}{2}\overline{s}(b^a *c_a)
 \nonumber \\ & =&  \frac{i}{2} \overline{s} s(h^{ab}*h_{ab})- \frac{i\alpha}{2}s(b^a *\overline{c}_a),
\end{eqnarray}
where  $s$ denotes the BRST transformations which is given by 
\begin{eqnarray}
s \,h_{ab} &=& \partial_a c_b + \partial_b c_a + \pounds_{(c)} h_{ab}, \nonumber \\
s \,c^a &=& - c_b *\partial^b c^a, \nonumber \\
s \,\overline{c}^a &=& b^a, \nonumber \\ 
s \,b^a &=&0,
\end{eqnarray}
and $\overline{s}$ denotes the anti-BRST transformations which is  given by 
\begin{eqnarray}
\overline{s} \,h_{ab} &=& \partial_a \overline{c}_b + \partial_b \overline{c}_a + \pounds_{(\overline{c})} h_{ab}, \nonumber \\
\overline{s} \,c^a &=& -b^a - 2 \overline{c}_b *\partial^b c^a, \nonumber \\
\overline{s} \,\overline{c}^a &=& - \overline{c}_b *\partial^b \overline{c}^a,\nonumber \\ 
\overline{s} \,b^a &=& - b^b*\partial_b c^a.
\end{eqnarray}
Here $ \pounds_{(c)}$ and $\pounds_{(\overline{c})}$ are given by 
\begin{eqnarray}
 \pounds_{(c)} h_{ab} = c^c *\partial_c h_{ab} + h_{ac}*\partial_b c^c + h_{cb} *\partial_a c^c, \nonumber \\ 
\pounds_{(\overline{c})} h_{ab} = \overline{c}^c* \partial_c h_{ab} + h_{ac}*\partial_b \overline{c}^c + h_{cb} *\partial_a \overline{c}^c.
\end{eqnarray}
Now using $c^a * c_a =  \overline{c}^a * \overline{c}_a =0$
and the Bianchi indentity, we can show after a 
straightforward calculation that 
these transformations are nilpotent,
\begin{equation}
 s^2 = \overline{s}^2 =  0.
\end{equation}
In fact, they can also be shown to satisfy  $\overline{s}s + s \overline{s} =0$.
Now the effective Lagrangian density which is given by the sum of the original classical Lagrangian density with these
gauge fixing and ghost terms is invariant under these noncommutative BRST and noncommutative anti-BRST transformations.
\begin{equation}
s\,  \mathcal{L}_{eff} = \overline{s} \, \mathcal{L}_{eff}  = 0,
\end{equation}
where 
\begin{equation}
\mathcal{L}_{eff} = \mathcal{L}_c + \mathcal{L}_g.
\end{equation}
It is so because for the original classical Lagrangian the noncommutative BRST or the anti-BRST noncommutative transformations are only 
gauge transformations with the gauge parameter replaced by ghosts or the anti-ghosts. Furthermore, as the sum of the 
gauge fixing and the ghost term can be written as a total noncommutative BRST or a total noncommutative anti-BRST  variation and both these 
transformations are nilpotent, the Lagrangian is invariant under them.  
It is obvious the in Landau gauge as $\alpha =0$, we can express the gauge fixing Lagrangian density as a 
combination of total noncommutative BRST and total noncommutative anti-BRST variations
\begin{eqnarray}
\mathcal{L}_g &=&-\frac{i}{2}s \overline{s} (h^{ab}*h_{ab}) \nonumber \\ & =&  \frac{i}{2} \overline{s} s(h^{ab}*h_{ab}).
\end{eqnarray}

\section{ Non-Linear Gauges} 
It is known that for  perturbative quantum gravity in Curci-Ferrari gauge
 we can write the sum of the gauge fixing term and the ghost term   as a combination of total BRST and total
 anti-BRST variations for any value of $\alpha$ \cite{1a}. We will show here that this can also be done for noncommutative perturbative 
quantum gravity. 
The noncommutative BRST transformations  for  noncommutative perturbative quantum gravity  in Curci-Ferrari gauge can be written as 
\begin{eqnarray}
s \,h_{ab} &=& \partial_a c_b + \partial_b c_a + \pounds_{(c)} h_{ab}, \nonumber \\
s \,c^a &=& - c_b* \partial^b c^a, \nonumber \\
s \,\overline{c}^a &=& b^a - \overline{c}^b*\partial_b c^a, \nonumber \\ 
s \,b^a &=& - b^b*\partial_b c^a -  \overline{c}^b*\partial_b c^d*\partial_d c^a,
\end{eqnarray}
and the   noncommutative anti-BRST transformation for noncommutative perturbative quantum gravity in Curci-Ferrari gauge can be written as 
\begin{eqnarray}
\overline{s}\, h_{ab} &=& \partial_a \overline{c}_b + \partial_b \overline{c}_a + \pounds_{(\overline{c})} h_{ab}, \nonumber \\
\overline{s} \,\overline{c}^a &=& - \overline{c}_b* \partial^b \overline{c}^a, \nonumber \\
\overline{s} \,c^a &=& - b^a - \overline{c}^b*\partial_b c^a, \nonumber \\ 
\overline{s} \,b^a &=& - b^b* \partial_b \overline{c}^a +  c^b* \partial_b\overline{c}^d*\partial_d  \overline{c}^a.
\end{eqnarray}
Now using $c^a * c_a =  \overline{c}^a * \overline{c}_a =0$
and the Bianchi indentity, we can show after a 
straightforward but lengthy calculation that 
these transformations are nilpotent,
\begin{equation}
 s^2 = \overline{s}^2 = 0.
\end{equation}
In fact, they also satisfy $ \overline{s}s + s \overline{s} =0$. 
We can now write a gauge fixing term and the ghost term  as a combination of a total BRST and a total  anti-BRST variation, as
\begin{eqnarray}
\mathcal{L}_g&=& \frac{i}{2}s\overline{s}\left[h^{ab}*h_{ab} - i \alpha \overline{c}^a* c_a \right] 
\nonumber \\ &=&\frac{-i}{2}\overline{s} s \left[h^{ab}*h_{ab} - i \alpha \overline{c}^a *c_a \right].
\end{eqnarray}

Now we can analyse the effect of the addition of a bare mass term to this Lagrangian density. The 
Lagrangian density for the  noncommutative perturbative quantum gravity in massive Curci-Ferrari  gauge can be written as  
\begin{eqnarray}
\mathcal{L}_{g}&=& \frac{i}{2}[s\overline{s}-im^2]\left[h^{ab}*h_{ab} - i \alpha \overline{c}^a* c_a \right] 
\nonumber \\ &=&\frac{i}{2}[-\overline{s} s-im^2] \left[h^{ab}*h_{ab} - i \alpha \overline{c}^a *c_a \right],
\end{eqnarray}
where the noncommutative  BRST transformations are given by 
\begin{eqnarray}
s \,h_{ab} &=& \partial_a c_b + \partial_b c_a + \pounds_{(c)} h_{ab}, \nonumber \\
s \,c^a &=& - c_b* \partial^b c^a, \nonumber \\
s \,\overline{c}^a &=& b^a - \overline{c}^b*\partial_b c^a, \nonumber \\ 
s \,b^a &=& i m^2 c^a- b^b*\partial_b c^a -   \overline{c}^b*\partial_b c^d* \partial_d c^a,
\end{eqnarray}
and the noncommutative  anti-BRST transformations are given by
\begin{eqnarray}
\overline{s}\, h_{ab} &=& \partial_a \overline{c}_b + \partial_b \overline{c}_a + \pounds_{(\overline{c})} h_{ab}, \nonumber \\
\overline{s} \,\overline{c}^a &=& - \overline{c}_b *\partial^b \overline{c}^a, \nonumber \\
\overline{s} \,c^a &=& - b^a - \overline{c}^b*\partial_b c^a, \nonumber \\ 
\overline{s} \,b^a &=& i m^2 \overline{c}^a- b^b*\partial_b  \overline{c}^a +  c^b*\partial_b\overline{c}^d*\partial_d  \overline{c}^a.
\end{eqnarray}

The addition of bare mass term breaks the nilpotency of these noncommutative  BRST and  anti-BRST transformations. 
 The BRST and the anti-BRST transformations now satisfy
\begin{equation}
  s^2 = \overline{s}^2  \sim i m^2.
\end{equation}
However,  the nilpotency of the BRST and the anti-BRST transformations is restored in the zero mass limit. 

It may be noted that the Lagrangian density for sum of the ghost term and gauge fixing term in
  the Landau gauge and these non-linear gauges can be expressed as a 
combination of total BRST and total anti-BRST variations. So the effective Lagrangian density in these gauges 
is also invariant under a symmetry called 
  the $FP$-conjugation in these gauges. This $FP$-conjugation is given by 
\begin{eqnarray}
\delta_{FP} \,h_{ab} &=&0, \nonumber \\
\delta_{FP}\,c^a &=& \overline{c}^a, \nonumber \\
\delta_{FP} \,\overline{c}^a &=& - c^a, \nonumber \\ 
\delta_{FP} \,b^a &=&b^a -2 c^b*\partial_b c^a.
\end{eqnarray}
 Even though the nilpotency is violated in the massive Curci-Ferrari gauge,
 the $FP$-conjugation is maintained.
The invariance of ordinary Yang-Mills 
theories under $FP$-conjugation for these gauges 
is well known \cite{y1,y2,y3,y4}. 

\section{Conserved Charges}
Conserved charges can only be defined properly for spacelike noncommutativity. This is 
because if we impose the full spacetime noncommutativity, we will end up having 
higher order time derivatives in out theory. This will spoil the unitarity of the theory.
Thus, we will now restricted our discussions to
spacelike noncommutativity. So, we  will set $\theta^{i0} =0$. 
In ordinary field theories there exists a divergenceless current corresponding to 
each symmetry transformation. 
In noncommutative field theories the 
divergence of the corresponding current  does not vanish. 
It is rather  equal to the Moyal bracket of some functions  \cite{msb}. However, this
Moyal bracket vanishes  for the spacelike noncommutativity,  when it is integrated 
over all spatial coordinates \cite{msb1}. 
Hence, again a conserved charge can be associated with a symmetry transformation. 
This conserved charge commutes with the Hamiltonian of the theory. 
Let $f(x)$ and $g(x)$ be two local functions in a noncommutative spacetime.  Now  the divergence of 
the current associated with a symmetry in this spacetime, can written as  
\begin{equation}
 [f(x), g(x)]_{*} = \partial^a J_a, 
\end{equation}
where we have defined the bracket $[f(x), g(x)]_{*}$ as 
\begin{equation}
 [f(x), g(x)]_{*} = f(x)* g(x) - g(x)* f(x). 
\end{equation}
In the case of spacelike noncommutativity, we have 
 $\theta^{i0} =0$, and so we get 
\begin{equation}
 \int d^3x [f(x), g(x)]_{*} =0.
\end{equation}
So,  we can write the  conserved charge corresponding to the symmetry as
\begin{equation}
Q = \int d^3 x    J^0.
\end{equation}

The total Lagrangian which is given by the sum of the original Lagrangian of 
noncommutative perturbative quantum gravity, the gauge fixing 
term and the ghost term is invariant under the noncommutative BRST and the noncommutative
 anti-BRST transformations.
We can calculate the conserved charges corresponding to the invariance 
of this total Lagrangian of noncommutative perturbative quantum gravity under these transformations. To do so, we first calculate the currents 
associated with these transformations, 
\begin{eqnarray}
J^{(B)}_a(x) & = &\frac{ \partial \mathcal{L}_{eff}  }{\partial h_{ab} } * s\, h_{ab} +
 \frac{ \partial \mathcal{L}_{eff}  }{\partial c_{a} } * s\, c_{a} + \nonumber \\&&
\frac{ \partial \mathcal{L}_{eff}  }{\partial \overline{c}_{a} } * s\, \overline{c}_{a} + 
\frac{ \partial \mathcal{L}_{eff}  }{\partial b_{a} } *  s\, b_{a},\nonumber \\
 \overline{J}^{(B)}_a (x) & = &\frac{ \partial \mathcal{L}_{eff}  }{\partial h_{ab} } *  \overline{s}\, h_{ab} +
 \frac{ \partial \mathcal{L}_{eff}  }{\partial c_{a} } * \overline{s}\, c_{a} + \nonumber \\&&
\frac{\partial \mathcal{L}_{eff}  }{\partial \overline{c}_{a} } *\overline{s}\, \overline{c}_{a} + 
\frac{\partial \mathcal{L}_{eff}  }{\partial b_{a} } * \overline{s}\, b_{a}.
\end{eqnarray}
Here $J_a$ is the current associated with the noncommutative  BRST symmetry and $\overline{J}_a$ is the current 
associated with 
noncommutative anti-BRST symmetry. 

Now we can calculate  the BRST charge  $Q_B$ and anti-BRST charge $\overline{Q}_B$ associated 
with the currents $J_a$ and 
$\overline{J}$ as follows,  
\begin{eqnarray}
 Q_B &=& \int d^3 x    J_{(B)}^0, \nonumber \\ 
 \overline{Q}_B &=& \int d^3 x    \overline{J}_{(B)}^0.
\end{eqnarray}
 We can also define a conserved current $J^{(FP)}_{a}$ corresponding to $FP$-conjugation as
\begin{eqnarray}
J^{(FP)}_a (x) & = &\frac{ \partial \mathcal{L}_{eff}  }{\partial h_{ab} } * \delta_{FP}\, h_{ab} +
 \frac{ \partial \mathcal{L}_{eff}  }{\partial c_{a} } *\delta_{FP}\, c_{a} + \nonumber \\&&
\frac{ \partial \mathcal{L}_{eff}  }{\partial \overline{c}_{a} } * \delta_{FP}\, \overline{c}_{a} + 
\frac{ \partial \mathcal{L}_{eff}  }{\partial b_{a} } *  \delta_{FP}\, b_{a},
\end{eqnarray}
and the conserved charge corresponding to it as 
\begin{equation}
 Q_{FP} = \int d^3 x J_{(FP)}^0.
\end{equation}

In noncommutative perturbative gravity the BRST  and the anti-BRST charges 
are nilpotent for all gauges except the   massive Curci-Ferrari gauge.  
So, we  will first  restrict our discussion to gauges other then the 
 massive Curci-Ferrari gauge. Then, we will analyse the effect of having 
the massive Curci-Ferrari gauge. 
As we have restricted our discussion to gauges other than the massive Curci-Ferrari gauge, 
so for any state 
 $|\phi\rangle$,  we have 
\begin{eqnarray}
 Q_B^2 |\phi\rangle &=& 0, \nonumber \\ 
 \overline{Q}_B^2 |\phi\rangle &=& 0.
\end{eqnarray}
We now define the  physical states  $ |\phi_p \rangle $
 as the states  annihilated by the  BRST charge
\begin{equation}
 Q_B |\phi_p \rangle =0.   
\end{equation}
We will obtain the  same result 
if we define the physical states as the states  annihilated by the anti-BRST charge
 \begin{equation}
 \overline{Q}_B |\phi_p \rangle =0. 
\end{equation}
We will 
get the same physical result by using either of these definitions of the physical states. 
The physical states that are obtained from other states by the action of 
either the BRST or the anti-BRST charges, 
are orthogonal to all  physical states. They are even
orthogonal to themselves. 
Thus,  two physical states  that differ from each other by 
such a state   will be indistinguishable.  
Let the asymptotic physical states be 
\begin{eqnarray}
 |\phi_{pa,out}\rangle &=& |\phi_{pa}, t \to \infty\rangle, \nonumber \\
 |\phi_{pb,in}\rangle &=& |\phi_{pb}, t \to - \infty\rangle.
\end{eqnarray}
 Now  a $\mathcal{S}$-matrix element can be written as
\begin{equation}
\langle\phi_{pa,out}|\phi_{pb,in}\rangle
 = \langle \phi_{pa}|\mathcal{S}^{\dagger}\mathcal{S}|\phi_{pb}\rangle.
\end{equation}
The BRST and the  anti-BRST charges  
 commute with the Hamiltonian because they are conserved charges. So, 
 the time evolution of any physical state will 
also be annihilated by  them. 
\begin{eqnarray}
 Q_B \mathcal{S} |\phi_{pb}\rangle &=&0, \nonumber \\
\overline Q_B \mathcal{S} |\phi_{pb}\rangle &=&0
\end{eqnarray}  
Thus, the states $\mathcal{S}|\phi_{pb}\rangle$ 
can only be a linear combination of physical states,
\begin{equation}
\langle\phi_{pa}|\mathcal{S}^{\dagger}\mathcal{S}|\phi_{pb}\rangle 
= \sum_{i}\langle\phi_{pa}|\mathcal{S}^{\dagger}|\phi_{0,i}\rangle
\langle\phi_{0,i}| \mathcal{S}|\phi_{pb}\rangle.
\end{equation}
Since the full $\mathcal{S}$-matrix is unitary this relation implies that the 
 $S$-matrix restricted to physical sub-space is also unitarity. 
It may be noted that the nilpotency of the BRST and the anti-BRST   charges was essential for 
proving the unitarity of the  $\mathcal{S}$-matrix. Now as the  BRST 
and the  anti-BRST 
charges are not nilpotent  in the massive Curci-Ferrari  gauge,
\begin{eqnarray}
 Q_B^2 |\phi\rangle &\neq& 0, \nonumber \\ 
 \overline{Q}_B^2 |\phi\rangle &\neq& 0.
\end{eqnarray}
So the above argument does not hold for the massive Curci-Ferrari  gauge. 
Thus,  the $\mathcal{S}$ does not factorize in the massive Curci-Ferrari  gauge
\begin{equation}
 \langle\phi_{pa}|\mathcal{S}^{\dagger}\mathcal{S}|\phi_{pb}\rangle 
\neq \sum_{i}\langle \phi_{pa}|\mathcal{S}^{\dagger}|\phi_{0,i}\rangle
\langle\phi_{0,i}| \mathcal{S}|\phi_{pb}\rangle,
\end{equation}
and  the resultant theory is not unitarity.
However, 
 the nilpotency  of the BRST and the anti-BRST charges is restored 
in the zero mass limit. As the unitarity of the theory depended 
on the nilpotency of these charges, so the unitarity is also restored in the zero mass limit. 
This loss of unitarity can have important 
physical consequences as it is expected that certain quantum gravitational process might 
lead to the violation of unitarity \cite{lqc}.

\section{Conclusion}

In this paper we have  generalized certain results of  perturbative quantum gravity
to noncommutative spacetime.     We  have expressed 
the gauge fixing Lagrangian density for perturbative quantum gravity as a
 combination of  total BRST and total anti-BRST variations in Landau gauge and non-linear gauges.
We  have shown that the nilpotency of the noncommutative BRST and the noncommutative anti-BRST leads to  the unitarity of  theory.  
Furthermore, as the addition of a bare mass term  
violated the nilpotency of the noncommutative BRST and the noncommutative anti-BRST transformations, so the  unitarity of 
 noncommutative perturbative quantum gravity is also violated in the  massive Curci-Ferrari  gauge. 
This may have important physical consequences for those processes in quantum gravity where the unitarity could be violated \cite{lqc}. 
 All these results were  already known to hold for 
commutative  perturbative quantum gravity  and we have shown here  that they 
also hold for noncommutative perturbative quantum gravity. 

The BRST and the anti-BRST symmetries for the noncommutative Yang-Mills theories have only been studied in linear gauges. 
So it will be interesting to analyse the BRST and the 
anti-BRST symmetries for the noncommutative Yang-Mills theories in non-linear gauges.  
It is known for ordinary Yang-Mills theories in the Landau and Curci-Ferrari gauges  that the BRST
and the anti-BRST are generators  are part of a larger $SL(2,R)$  algebra
known as the Nakanishi-Ojima algebra \cite{y1}. It will be interesting to find out if  this algebra also holds 
for noncommutative Yang-Mills 
theories and noncommutative perturbative quantum gravity. 
 Furthermore, it is known  for ordinary Yang-Mills theories that this algebra is broken by ghost condensation \cite{y4}.  It 
is highly likely that a similar thing occurs in noncommutative Yang-Mills theories. However, the study of ghost condensation for 
noncommutative perturbative quantum gravity might be highly non trivial. This is so become 
 ghost condensation  can generate a vector field with non-vanishing 
vacuum expectation value. This can break the Lorentz symmetry. 
 The ghost  condensation is also expected to 
 modify the infrared behavior of the off-diagonal ghost
propagator, while contributing to the vacuum energy density. 

It will also be interesting to generalise the results of this paper to general curved spacetime.
The generalisation to arbitrary spacetimes might not be  simple as it is still not  completely clear
 how BRST symmetry will work their. However, the generalisation to Anti-de Sitter  spacetime 
might be  straightforward because the ghost fields are expected to be well behaved in Anti-de Sitter  spacetime \cite{fa}.
 For de Sitter spacetime their are issues with 
 BRST invariance of the theory   due to the infrared divergence of the ghost 
propagators \cite{fa}. Thus the generalisation of this work to de Sitter spacetime   will not be straightforward. 

It will also be useful to
analyse the shift symmetry for these noncommutative theories in the anti-field formalism and express the result in 
superspace formalism. As the effect of shift symmetry 
in commutative field theories with higher derivatives has already been analysed \cite{fm},  
it will be interesting to generalise those result to noncommutative field theories including 
noncommutative perturbative quantum gravity.

\end{document}